\documentclass[showpacs,twocolumn,prb]{revtex4-1}
\usepackage{graphicx}
\newcommand{\be}{\begin{equation}}
\newcommand{\ee}{\end{equation}}
\newcommand{\bea}{\begin{eqnarray}}
\newcommand{\eea}{\end{eqnarray}}
\newcommand{\bwt}{\begin{widetext}}
\newcommand{\ewt}{\end{widetext}}

\begin{document}

\title{The polaron-like nature of an electron coupled to phonons}
\author{Zhou Li and F. Marsiglio}
\affiliation{Department of Physics, University of Alberta, Edmonton, Alberta,
Canada, T6G~2E1}

\begin{abstract}
When an electron interacts with phonons, the electron can exhibit either free electron-like or polaron-like properties. The latter tends to occur for very strong coupling, and results in a phonon cloud accompanying the electron as it moves, thus raising its mass considerably. We summarize this behaviour for the Holstein model in one, two and three dimensions, and note that the crossover occurs for fairly low coupling strengths compared to those attributed to real materials exhibiting conventional superconductivity.
\end{abstract}

\date{\today}
\maketitle

\section{Introduction}
The conventional theory of superconductivity consists of two distinct pieces. The first describes the formalism of pairing, premised on the idea that two electrons effectively attract one another; this pairing is well described by the Bardeen-Cooper-Schrieffer (BCS) wave function,\cite{bardeen57} and continues to be used extensively to describe almost all known superconductors. The second piece concerns the origin of the attractive interaction, suggested originally by 
Fr\"ohlich \cite{frohlich50} and Bardeen\cite{bardeen50} to be the electron-phonon interaction. In the BCS calculation the  additional piece concerning the origin (i.e. {\em mechanism}) of pairing enters only in that a simple (though confusing) cutoff of order the Debye (phonon) frequency is included in the attractive interaction. It is confusing because it is added in momentum space, whereas the physical cutoff is in frequency space, to reflect the retardation effects expected from relatively sluggish phonons interacting with nimble electrons. The cutoff occurs naturally in frequency space in the later formulation of Eliashberg\cite{eliashberg60} and others.\cite{nambu60,morel62,schrieffer63} 

A very solid and coherent case has been made for this description of phonon-mediated mechanism of superconductivity
for simple metals like Lead.\cite{parks69,allen82} However, even though Eliashberg theory is often referred to as a microscopic theory of superconductivity (and, indeed, it is {\em more} microscopic than BCS theory), it retains some elements of a phenomenology. This occurs in two distinct areas in the theory; the first concerns the treatment of direct Coulomb interactions, which in both BCS and Eliashberg treatments tends to be modelled through the use of a single (renormalized) parameter, $\mu^\ast$, and the second concerns the treatment of the phonons. Traditionally, the phonons are taken from experiment, notably tunneling\cite{mcmillan69} or neutron scattering.\cite{brockhouse62} In either case all the ``calculations'' concerning the phonons have been done by nature, and handed over to the theorist to incorporate into Eliashberg theory. Notable exceptions are, for example, treatments in Ref. (\onlinecite{marsiglio90}) and more recently in Ref. (\onlinecite{bauer11}), where an attempt is made to compute modification to the properties of the phonons alongside those of the electrons, due to strong electron-phonon interactions.

As an example we show in Fig.~1 a plot of the pairing susceptibility (a measure of the superconducting critical temperature) vs. electron filling, using a Migdal-Eliashberg calculation with and without phonon renormalization.\cite{marsiglio91} When phonon renormalization is included, the susceptibility (here plotted for an inverse temperature, 
$\beta = 6$), decreases, particularly as one approaches half filling. In fact our calculations stop because a charge density wave instability occurs for a density less than half filling. This makes apparent that phonon renormalization has an important but deleterious effect on superconductivity, at least for the model considered in Ref. (\onlinecite{marsiglio91}). This is an important point which will have to be reconsidered in future work. For the present, however, we wish to focus on a simpler model, that of a {\em single} electron interacting with many phonons. This object is usually referred to as a polaron (though not in the literature pertaining to superconductivity\cite{remark1}), and we regard the polaron as the basic building block, from which a theory of interacting electrons and phonons must be built. This is {\em not} how the theory of superconductivity progressed historically. A weak coupling approach was assumed at the onset; this means that a (large) Fermi sea is taken as `given', and then Migdal-like\cite{migdal58} arguments are adopted to essentially use weak coupling perturbation theory.\cite{remark2}

\begin{figure}
\begin{center}
\includegraphics[height=3.3in,width=2.5in, angle = -90]{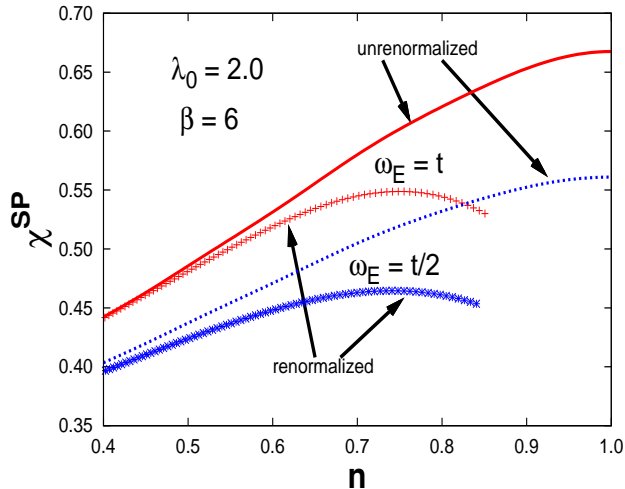}
\caption{(Color online)
Pairing susceptibility vs. electron filling for a fixed bare coupling strength, $\lambda_0 = 2$, and a fixed temperature, given by $\beta \equiv 1/(k_B T) = 6$. Results for two different frequencies are shown; usual Migdal calculations of the electronic properties with no changes to the phonons are given by the curves, while results that allow the phonons to renormalize due to the electron phonon coupling are given by the symbols. Clearly allowing for phonon renormalization leads to a suppression of the pairing susceptibility, and therefore superconductivity.}
\end{center}
\end{figure}

In this work we wish to illustrate the polaronic nature of the electron when coupled to phonons, using the Holstein model.\cite{holstein59} We will summarize results in one, two, and three dimensions, adopting various strategies: weak and strong coupling perturbation theory,\cite{marsiglio95,li10} the adiabatic approximation,\cite{kabanov93} and an exact diagonalization method which incorporates Bloch's Theorem which we refer to as the Trugman method.\cite{trugman90,bonca99,ku02} Very little work has been done beyond the Holstein model. When a non-local interaction is used, the coupling in momentum space acquires momentum dependence, making all techniques more difficult. Very recently we have succeeded in obtaining analytical results for the polaron in the weak coupling limit\cite{li11} of the 
Bari\u si\'c, Labb\' e, and Friedel\cite{barisic70}-Su-Schrieffer-Heeger\cite{su79} (BLF-SSH) model, which models acoustic phonons coupled to electrons through their itineracy. Other, more complicated models have also been proposed. For example in Ref. (\onlinecite{hague06}) a longer range electron phonon interaction is proposed, and Jahn-Teller-type distortions in the presence of repulsive Coulomb interactions also lead to polaron formation.\cite{mertelj07} Very recently, Innocenti {\it et al.}\cite{innocenti10} have proposed a two electron band scenario, where tuning of the chemical potential can lead to polaron formation; a companion proposal\cite{poccia11} also suggests that so-called `dopant self-organization' can lead to trapping of polarons, resulting in some kind of inhomogeneous phase.
These are all interesting possibilities, but here we focus on rigorous results for the simplest model describing polarons.

In the next section we describe the various techniques used to attack the Holstein polaron, followed by results for two specific properties --- the ground state energy and the effective mass. It is seen, for example, that the electron is always polaronic in one dimension. More precisely, this statement means the following: as the phonon frequency is taken to be smaller and smaller (this is necessarily a process taken as a limit) more and more phonons are required to properly describe the ground state, irrespective of the coupling strength. In the past researchers have referred to this as a self-localized (or self-trapped) state; this is a misnomer, as no such violation of Bloch's Theorem occurs. Partly this nomenclature has arisen from variational studies (and the adiabatic limit, where the phonon frequency $\omega_E$ is taken to be zero at the onset), where then a localized state does arise. The result is that the entity of interest emerging out of the ground state is hardly recognizable as an electron; rather it contains a considerable admixture of phonons, and hence is renamed a polaron. After presenting results in two and three dimensions, where a regime of weak coupling exists in which the electron definitely retains free-electron-like ( and therefore `non-polaron-like') character for weak coupling strengths, we will conclude with a summary.

\section{Methods}

Perhaps the simplest starting point for all studies such as this one is the weak coupling approximation. This is well-defined with straightforward perturbative methods, or one can adopt the language of Green functions, where the electron propagator, $G(k,\omega)$, can be written in momentum ($k$) and frequency ($\omega$) space in terms of a self energy, $\Sigma(k,\omega)$:
\be
\Sigma(k,\omega+ i\delta)  =  -\sum_{k^\prime} |g(k,k^\prime)|^2 G_0(k^\prime,\omega + i\delta - \omega(k-k^\prime)),
\label{selfenergy}
\ee
where $G_0(k,\omega+i\delta) \equiv \bigl[ \omega + i\delta - \epsilon_k \bigr]^{-1}$ is the non-interacting electron retarded propagator, $\epsilon_k$ ($\omega_q$) gives the dispersion relation for the non-interacting electrons (phonons),  respectively, and $g(k,k^\prime)$ describes any possible momentum dependence in the electron-phonon coupling. Note that the infinitesimal imaginary part to the frequency, $i\delta$, is necessary for the analyticity of the Green function; here we use the retarded functions (positive imaginary part). This is written quite generally, and for any dimension. Here, we wish to discuss specifically the Holstein model, with Hamiltonian written in momentum space as
\bea
H&=&\sum_k \epsilon_k c_k^\dagger c_k + \omega_E \sum_q a_q^\dagger a_q \nonumber \\
&-& {g\omega_E \over \sqrt{N}} \sum_{k,k^\prime} c_k^\dagger c_{k^\prime}(a_{k-k^\prime} + a_{-(k - k^\prime)}^\dagger),
\label{hol_ham}
\eea
where $\omega_E$ is the Einstein phonon frequency and $g$ is the (dimensionless) electron-phonon coupling, and in this case is clearly a constant. We will tend to use the (also dimensionless) parameter $\lambda$, which will have a different definition depending on dimension. In one dimension, we use $\lambda \equiv g^2\omega_E/(W/2) = g^2\omega_E/(2t) $, where $W \equiv 4t$ is the electron bandwidth written in terms of the nearest neighbour hopping parameter, $t$. Hereafter $t$ will be set to unity. 

\begin{figure*}[tp]
\centering
\includegraphics[height=2.8in,width=3.0in]{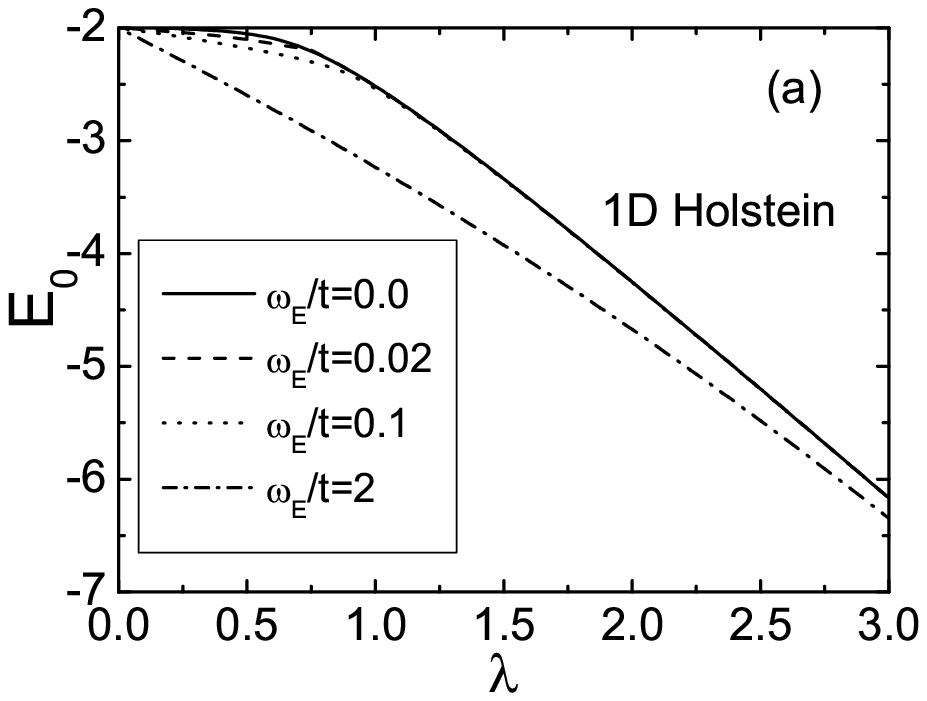}
\includegraphics[height=2.8in,width=3.0in]{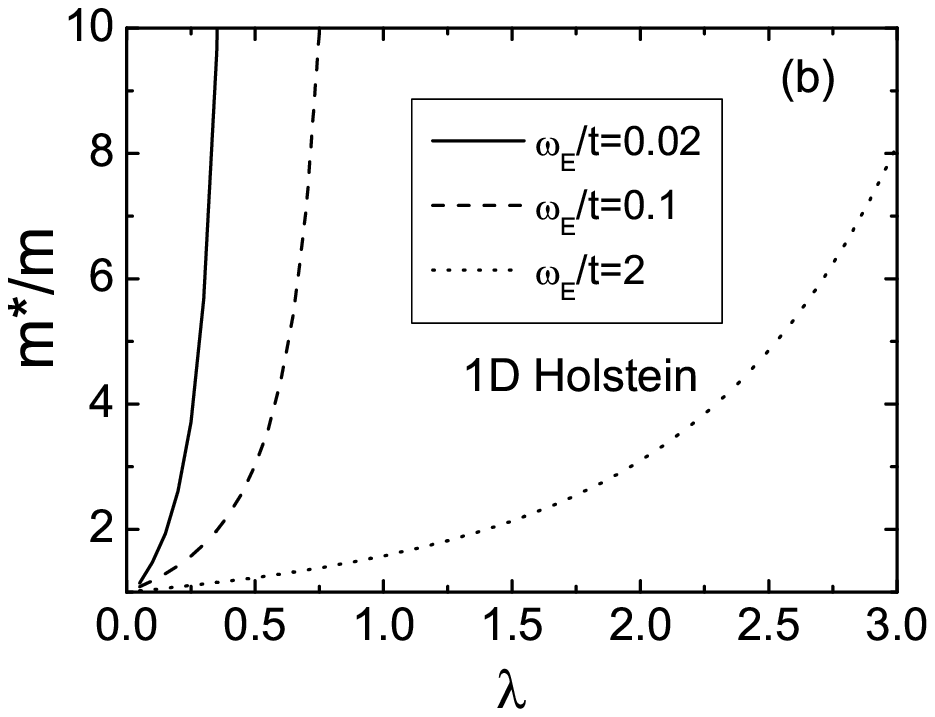}
\includegraphics[height=2.8in,width=3.0in]{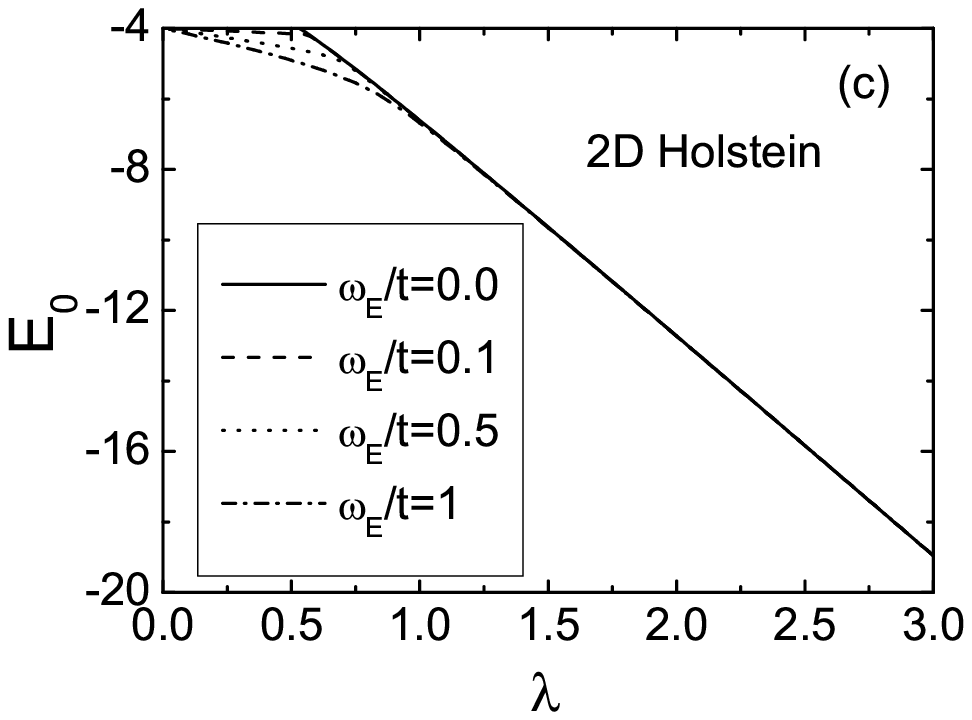}
\includegraphics[height=2.8in,width=3.0in]{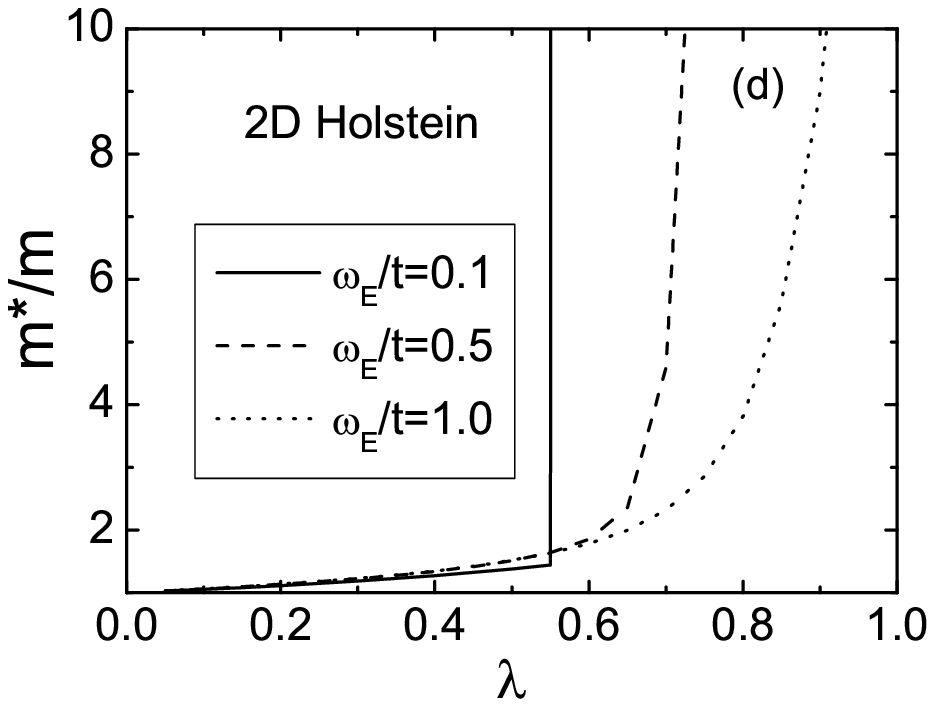}
\includegraphics[height=2.8in,width=3.0in]{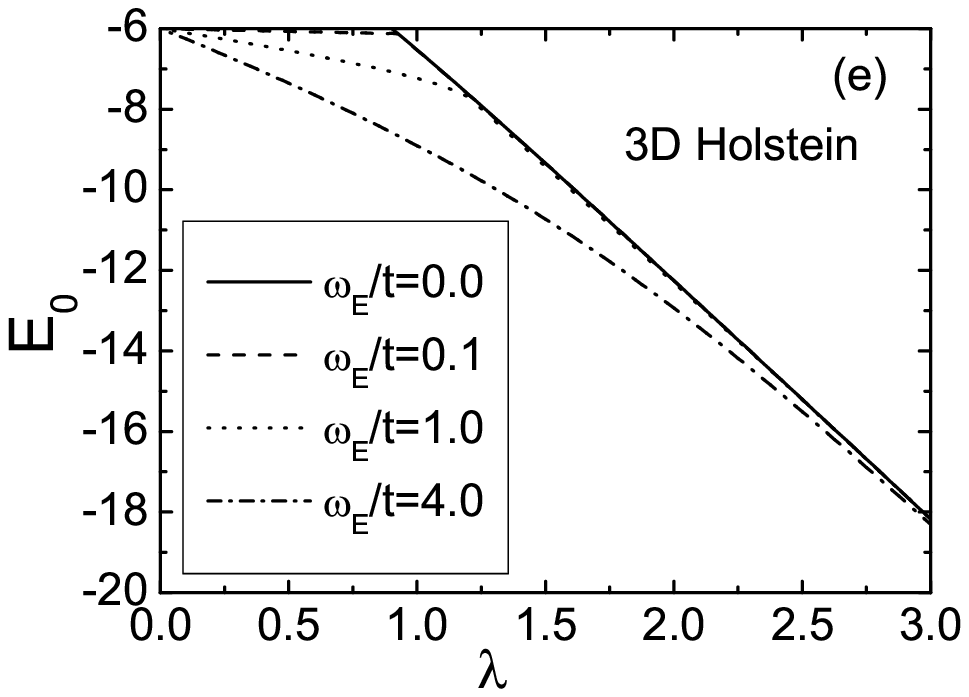}
\includegraphics[height=2.8in,width=3.0in]{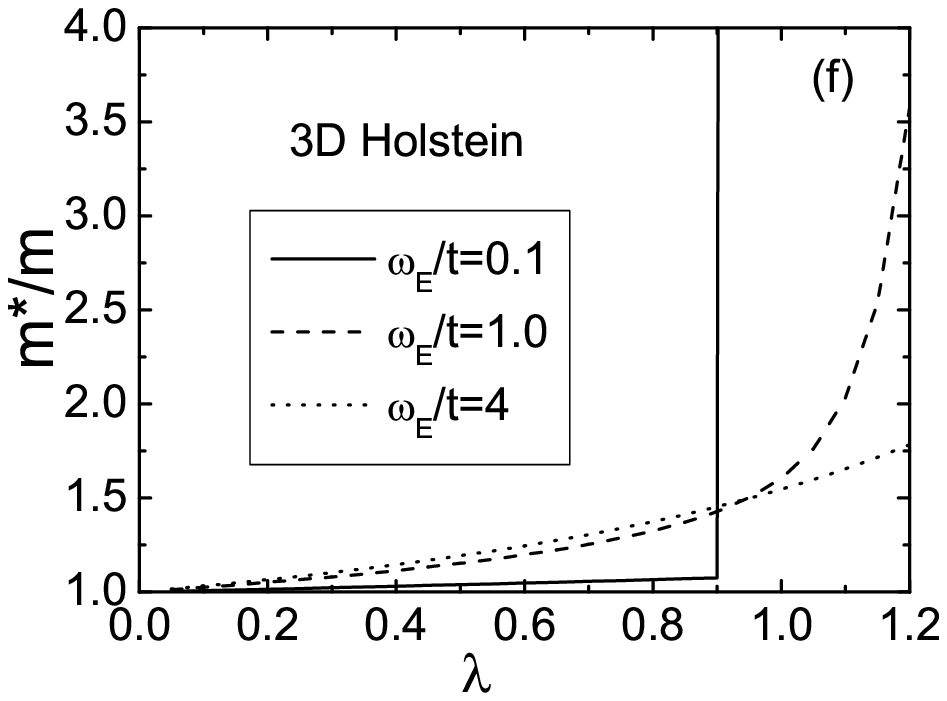}
\caption{The ground state energy and effective mass vs electron phonon coupling strength for the Holstein model. The energy ((a) (c) (e)) and effective mass ((b) (d) (f)) are plotted in one, two, and three dimensions, respectively. Note the very abrupt crossover to polaron-like behaviour in 2D and 3D, both at relatively low coupling strengths. The strong coupling limit agrees very well with the adiabatic limit in all three dimensions. Also note that in (a), (c) and (e) there are three curves that are indistinguishable for $\lambda {{ \atop {> \atop \sim}} \atop } 1.5$, showing that the strong coupling and adiabatic regime is readily achieved for moderate $\lambda$ and non-zero $\omega_E$.}
\label{fig2}
\end{figure*}

In one and two dimensions, the momentum sums can be evaluated analytically. The results are
\begin{equation}
\Sigma_{\mathrm{1D}}(\omega+i\delta) = {\frac{ \lambda \omega_E \mathrm{sgn}%
(\omega - \omega_E) }{\sqrt{\biggl({\frac{\omega - \omega_E }{2t}}\biggr)^2
- 1}}},
\label{sig_weak_1D}
\end{equation}
so that the ground state energy is
\begin{equation}
E_0 = -2t\biggl( 1 + \lambda \sqrt{\frac{\omega_E }{4t + \omega_E}} \biggr),
\label{e0_weak_1D}
\end{equation}
and the effective mass $m^\ast$ is given by
\begin{equation}
m/m^\ast = 1 - {\frac{\lambda }{2}} \sqrt{\frac{t }{\omega_E}} {\frac{1 + {\frac{%
\omega_E }{2t}} }{(1 + {\frac{\omega_E }{4t}})^{3/2}}}.
\label{eff_mass_1D}
\end{equation}

In two dimensions we obtain\cite{li10}
\begin{equation}
\Sigma _{\mathrm{2D}}(\omega+i\delta) = {\frac{ \lambda }{2}}{\frac{%
8t\omega_E }{\omega - \omega_E}} K\Biggl[ \biggl({\frac{4t }{\omega -
\omega_E}}\biggr)^2\Biggr],  \label{sig_weak_2D}
\end{equation}
where $K(x) \equiv \int_0^{\pi/2} d\theta {\frac{1 }{\sqrt{1 - x \sin^2{%
\theta}}}}$ is the complete Elliptic integral of the first kind. This leads
to a ground state energy, which, in weak coupling, is:
\begin{equation}
E_0 = -4t \Biggl( 1 + {\frac{\lambda }{4}}{\frac{\omega_E }{t}}{\frac{1 }{1
+ \omega_E/(4t)}} K\Biggl[ {\frac{1 }{\bigl(1 + \omega_E/(4t)\bigr)^2}} %
\Biggr] \Biggr).  \label{e0_weak_2D}
\end{equation}
We can take the derivative of Eq. (\ref{sig_weak_2D}) to obtain:
\begin{equation}
m^\ast/m = 1 + {\frac{\lambda }{2}} {\frac{1 }{1 + \omega_E/(8t)}} E\biggl[{1 \over
(1+\omega_E/(4t))^2}\biggr],  \label{mass_weak_2D}
\end{equation}
where $E(x) \equiv \int_0^{\pi/2} d\theta \sqrt{1 - x \sin^2{\theta}}$ is
the complete Elliptic integral of the second kind. 
Three dimensions is straightforward, but requires numerical integration.

The calculation in the strong coupling limit is outlined in Ref. (\onlinecite{marsiglio95}). 
The unperturbed state is:
\begin{equation}
|\psi \rangle = e^{-g^2/2} \sum_\ell e^{ikR_\ell} e^{-g\hat{a}_\ell^\dagger} \hat{c}%
_\ell^\dagger |0 \rangle,  \label{sc_zero}
\end{equation}
where the sum is over all lattice sites.\cite{remark3} Straightforward application of perturbation theory yields
a dispersion, albeit with exponentially suppressed bandwidth. This constitutes an estimate for the effective mass, which is far too high, and will be omitted from the plots. However, second order perturbation theory gives a substantive correction to the ground state energies; these are given, for large coupling values, as
\be
E_0 \approx -\omega_E g^2 - {dt^2 \over \omega_E g^2},
\label{strong_energy}
\ee
where $d$ is the dimension of the space. In the parts of Fig.~2 showing the ground state energy, the curve displaying Eq. (\ref{strong_energy}) is indistinguishable from the adiabatic limit for $\lambda {{ \atop {> \atop \sim}} \atop } 1$.
Adiabatic limit results are determined through a decoupling procedure, whereby the ion displacements are treated as
c-numbers.\cite{kabanov93,marsiglio95} 

Finally, exact results are obtained using the Trugman method,\cite{bonca99,ku02} We are able to obtain converged results for very small values of the dimensionless parameter $\omega_E/t$ by using Eq. (\ref{sc_zero}) as a `seed' state for the state construction. Iteration is considerably improved as well by starting in the strong coupling limit, converging results there, and using these results as subsequent `seeds' as we span the coupling by lowering the coupling strength a little at a time.

\section{Results and Discussion}

Results are shown in Fig.~2 for the one dimensional (a) and (b), two dimensional (c) and (d), and three dimensional (e) and (f) ground state energy and effective mass, respectively, vs. electron phonon coupling strength, for a variety of phonon frequencies. Note that we use $\omega_E/t$ and $\lambda$ as the two variables to characterize the adiabaticity and coupling strength, respectively, of the electron phonon system. Here 
$\lambda \equiv \omega_E g^2/(W/2)$ in one and three dimensions, where $W$ is the bandwidth (here, for a tight binding model with nearest neighbour hopping only, $W = 4t$ in 1D and $W = 12 t$ in 3D. In two dimensions we defined $\lambda$ slightly differently,\cite{li10} as $\lambda \equiv \omega_E g^2/(2 \pi t)$; this is because the non-interacting electron density of states (EDOS) is a constant at the bottom of the band, and so we adopted this value to multiply the bare value, $\lambda_0$, rather than the average EDOS, which is what we did in 1D and 3D. With this definition, for example, the effective mass becomes $1 + \lambda/2$ in 2D in the weak coupling adiabatic limit.\cite{li10}

The main results can be summarized as follows: in 1D the electron is {\em always} polaron-like. Note in particular that as the phonon frequency decreases the effective mass increases dramatically (Fig. 2(b)), and essentially diverges in the adiabatic limit, even for infinitesimal coupling strength. The results for the ground state go smoothly over to the adiabatic limit, as is apparent from the disappearance of two of the finite frequency curves into the adiabatic curve in Fig. 2(a).
In 2D and 3D there are clear delineations of free electron-like behaviour at weak coupling and polaron-like behaviour at intermediate and strong coupling. However all the results are `smooth' for a non-zero phonon frequency --- only when an adiabatic calculation is performed (with the phonon frequency set equal to zero at the onset) does the curve display a kink (and therefore a transition). So no phase transition occurs as long as $\omega_E \ne 0$. The remarkable result is that the crossover to polaron-like behaviour occurs at such an intermediate coupling strength,\cite{alexandrov_proc} well below the value of coupling strength normally assigned to conventional superconductors (even in 2D, if the average EDOS is used in the definition of  $\lambda$, the crossover coupling strength only moves up to about $0.86$, just a little lower than the value in 3D). Ongoing work is investigating this crossover behaviour when acoustic modes are involved.\cite{li11}

\label{Acknowledgments}
This work was supported in part by the Natural Sciences and Engineering
Research Council of Canada (NSERC), by ICORE (Alberta), and by the Canadian
Institute for Advanced Research (CIFAR).


\begin{thebibliography}{1}

\bibitem{bardeen57}
J. Bardeen, L.N. Cooper and J.R. Schrieffer, Phys. Rev. {\bf 106}, 162 (1957);
Phys. Rev. {\bf 108}, 1175 (1957).

\bibitem{frohlich50} H. Fr\"ohlich, Phys. Rev. {\bf 79} 845 (1950).

\bibitem{bardeen50} J. Bardeen, Phys. Rev. {\bf 79} 167 (1950).

\bibitem{eliashberg60} G.M. Eliashberg, Zh. Eksperim. i Teor. Fiz.
{\bf 38} 966 (1960); Soviet Phys.  JETP {\bf 11} 696 (1960).

\bibitem{nambu60} Y. Nambu, Phys. Rev. {\bf 117} 648 (1960).

\bibitem{morel62}P. Morel and P.W. Anderson, Phys. Rev. {\bf 125} 1263 (1962).

\bibitem{schrieffer63} J.R. Schrieffer, D.J. Scalapino  and J.W. Wilkins,
 Phys. Rev. Lett. {\bf 10} 336 (1963); D.J. Scalapino, J.R. Schrieffer and J.W. Wilkins,
Phys. Rev. {\bf 148} 263 (1966).

\bibitem{parks69} See {\it Superconductivity}, edited by R.D. Parks
(Marcel Dekker, Inc., New York, 1969), particularly the articles written by D.J. Scalapino (p.449) and
W.L. McMillan and J.M. Rowell (p. 561).

\bibitem{allen82} More recent reviews are: P.B. Allen and B. Mitrovi\'{c},
In: {\it Solid State Physics}, edited by H. Ehrenreich, F.~Seitz,
and D. Turnbull (Academic, New York, 1982) Vol. 37, p.1. See also
D. Rainer, In: {\it Progress in Low Temperature Physics},
Vol. 10, edited by D.F. Brewer (North--Holland, 1986), p.371, 
J.P. Carbotte, Rev. Mod. Phys. {\bf 62} 1027 (1990), and
F. Marsiglio and J.P. Carbotte, {\em
Electron-Phonon Superconductivity}, in `Superconductivity: Conventional and Unconventional
Superconductors' edited by K.H. Bennemann and J.B. Ketterson (Springer-Verlag),
pp. 73-162 (2008).

\bibitem{mcmillan69} W.L. McMillan and J.M. Rowell, In: {\it Superconductivity},
edited by R.D. Parks (Marcel Dekker, Inc., New York, 1969)p. 561.

\bibitem{brockhouse62}
B.N. Brockhouse, T. Arase, G. Caglioti, K.R. Rao and A.D.B. Woods,
Phys. Rev. {\bf 128} 1099 (1962).

\bibitem{marsiglio90} F. Marsiglio, Phys. Rev. B {\bf 42} 2416 (1990).

\bibitem{bauer11} J. Bauer, J.E. Han, and O. Gunnarsson, arXiv:1105.2833, (2011).

 \bibitem{marsiglio91} See also F. Marsiglio, in: {\it Electron--Phonon Interaction in Oxide Superconductors},
edited by R. Baquero (World Scientific, Singapore, 1991) p.167.

\bibitem{remark1} There is a remarkable divergence in the literature concerning the polaron. Though many of
the same researchers pioneered work in both the polaron problem and electron-phonon driven superconductivity in the early and mid-1950's, the concept of the polaron was essentially erased from discussion of superconductivity. Indeed, in the entire treatise by Parks (Ref. (\onlinecite{parks69})) the word `polaron' appears but once --- in the article by Gladstone et al.\cite{gladstone69} where it is mentioned only to `declare' that it is irrelevant to superconductivity. A definitive monogram on the polaron appeared in 1962,\cite{kuper63} where superconductivity, though of fervent interest at that time, is likewise never discussed. In the last 30 years, `polarons' and `superconductivity' have experienced more overlap, notably due to the work of Alexandrov and coworkers.\cite{alexandrov95} 

\bibitem{gladstone69} G. Gladstone, M.A. Jensen and J.R. Schrieffer, in: {\it Superconductivity},
edited by R.D. Parks (Marcel Dekker, Inc., New York, 1969)p. 665.

\bibitem{kuper63} In: {\it Polarons and Excitons} edited by C.G. Kuper
and G.D. Whitfield (Oliver and Boyd, Edinburgh, 1963).

\bibitem{alexandrov95} A.S. Alexandrov and N. Mott, {\it Polarons and
Bipolarons} (World Scientific, Singapore, 1995).

\bibitem{migdal58} A.B. Migdal, Zh. Eksp. Teor. Fiz. {\bf 34}, 1438 (1958)
[Sov. Phys. JETP {\bf 7}, 996 (1958)].

\bibitem{remark2} Note that Migdal\cite{migdal58} actually pointed out the necessity of including phonon renormalization in his calculation.

\bibitem{holstein59} T. Holstein, Ann. Phys. {\bf 8}, 325, (1959; ibid. {\bf 8}, 343 (1959).

\bibitem{marsiglio95} F. Marsiglio, Physica C\textbf{244} 21, (1995).

\bibitem{li10} Zhou Li, D. Baillie, C. Blois, and F. Marsiglio, Phys. Rev. B\textbf{81}, 115114, (2010).

\bibitem{kabanov93} V.V. Kabanov and O.Y. Mashtakov, Phys. Rev. B\textbf{47}, 6060 (1993).

\bibitem{trugman90} S. A. Trugman, in Applications of Statistical and Field Theory Methods to Condensed Matter, 
edited by D. Baeriswyl, A.R. Bishop and J. Carmelo, Plenum Press, New York, 1990.

\bibitem{bonca99} J. Bon\v ca, S.A. Trugman, and I. Batist\' ic, Phys. Rev. B\textbf{60},1633 (1999).

\bibitem{ku02} L-C. Ku, S.A. Trugman, and J. Bon\v ca, Phys. Reb. B\textbf{65}, 174306 (2002).

\bibitem{li11} Zhou Li, C.J. Chandler, and F. Marsiglio, Phys. Rev. B\textbf{83}, 045104, (2011).

\bibitem{barisic70} S. Bari\u si\' c, J. Labb\' e, and J. Friedel, Phys. Rev. Lett.\textbf{25}, 919 (1970); S. Bari\u si\' c, Phys. Rev. B\textbf{5}, 932 (1972), S. Bari\u si\' c, Phys. Rev. B\textbf{5}, 941 (1972).

\bibitem{su79} W.P. Su, J.R. Schrieffer, and A.J. Heeger, Phys. Rev. Lett.\textbf{42}, 1698 (1979); Phys. Rev. B\textbf{22}, 2099 (1980).

\bibitem{hague06} J. P. Hague, P. E. Kornilovitch, A. S. Alexandrov, and J. H. Samson, Phys. Rev. B{\bf 73}, 054303 (2006);  J.P. Hague et al. Phys. Rev. Lett. {\bf 98}, 037002 (2007).

\bibitem{mertelj07} T. Mertelj, V.V. Kabanov, J. Miranda Mena, and D. Mihailovic, Phys. Rev. B{\bf 76}, 054523 (2007).

\bibitem{innocenti10} D. Innocenti, N. Poccia, A. Ricci, A. Valletta, S. Caprara, A. Perali, and A. Bianconi, Phys. Rev. B {\bf 82}, 184528 (2010).

\bibitem{poccia11} N. Poccia, M. Fratini, A. Ricci, G. Campi, L. Barba,
A. Vittorini-Orgeas, G. Bianconi, G. Aeppli and A. Bianconi,
Nature Materials {\bf 10}, 733 (2011).

\bibitem{remark3} There is a large degeneracy in the strong coupling limit, as the electron can be placed on any site
of the lattice, with an accompanying phonon `cloud'. This state is chosen to satisfy Bloch's Theorem in anticipation of calculations to first and second order in the hopping.

\bibitem{alexandrov_proc} As has been proclaimed by Alexandrov for several decades now --- see Ref. (\onlinecite{alexandrov95}) and references therein.

\end{thebibliography}
\end{document}